\documentclass[12pt]{article}
\usepackage{graphicx}
\usepackage{subfigure}
\usepackage{amsfonts}
\usepackage{amssymb,amsmath}
\usepackage{hyperref}

\newcommand{\bea}{\begin{eqnarray}}
\newcommand{\eea}{\end{eqnarray}}

\makeatletter
\@addtoreset{equation}{section}

\makeatother

\usepackage{multicol}
\usepackage[usenames,dvipsnames]{xcolor}
\definecolor{rosso}{cmyk}{0,1,1,0.3}
\definecolor{verde}{cmyk}{0.8,0,0.6,0.25}
\definecolor{bluc}{cmyk}{1,0.4,0,0.1}
\definecolor{blucc}{cmyk}{0.8,0.3,0,0}

\def\be{\begin{equation}}
\def\ee{\end{equation}}

\def\MP{M_{\rm Pl}}
\def\({\left(}
\def\){\right)}
\def\1{^{(1)}}
\def\2{^{(2)}}

\def\<{\langle}
\def\>{\rangle}

\begin{document}

\begin{titlepage}

\begin{flushright}
UT-13-13\\
\end{flushright}

\vskip 3cm

\begin{center}

{\Large \bf 
Gamma-ray line from radiative decay of\\
 gravitino dark matter
}

\vskip .5in

{
Seng Pei Liew
}

\vskip .3in

{\em
Department of Physics, University of Tokyo, Bunkyo-ku, Tokyo 113-0033, Japan \vspace{0.2cm}
}

\end{center}

\vskip .5in

\begin{abstract}

We study radiative decay of gravitino dark matter with trilinear R-parity violations. We show that the branching ratio of the decay of gravitino into monochromatic photon can be large enough to explain the observed gamma-ray line from the Galactic centre in the Fermi-LAT data without producing too much continuum gamma-ray and anti-proton flux. This scenario is realized when the mass of sfermions and the trilinear R-parity violating coupling are $O(1-10)$ TeV and $O(10^{-7}-10^{-6})$ respectively. 

\vskip.5in
\noindent

\end{abstract}

\end{titlepage}

\setcounter{page}{1}

\section{Introduction} 
\label{sec:introduction}

Recent studies on the four-year Fermi data have found excess of 130 GeV gamma-ray line from 
the Galactic Center (GC)~\cite{Bringmann:2012vr, Weniger:2012tx,Tempel:2012ey,Boyarsky:2012ca,Su:2012ft,Hektor:2012kc}. There have been many papers studying various possible explanations (instrumental effects~\cite{Hektor:2012ev,Finkbeiner:2012ez,Whiteson:2012hr,Whiteson:2013cs}, pulsar wind effects~\cite{Aharonian:2012cs} etc.) of this signal but it is most interesting if it is to be interpreted as a dark matter (DM) signature~\cite{Bringmann:2012ez,Weniger:2013dya}. If this interpretation is correct, the 130 GeV gamma-ray line is the long-awaited signature of non-gravitational interactions of particle DM.

In order to explain the 130 GeV gamma-ray line with particle DM, a rather large branching ratio of DM annihilating or decaying into monochromatic photon is required, i.e. $\rm Br(\rm DM \to \gamma) \gtrsim 0.01$~\cite{Buchmuller:2012rc,Cohen:2012me,Cholis:2012fb}. Otherwise, DM to fermion and gauge boson annihilation or decay channels would produce too much continuum gamma-ray and conceal the line signal. Moreover, anti-proton flux produced by these channels are constrained by cosmic-ray observations~\cite{Adriani:2010rc}. However, in many cases, the branching ratio of DM annihilating or decaying into photons is suppressed  because DM does not couple directly to photon~\cite{Kyae:2012vi,Buckley:2012ws}. It has been shown that for monochromatic photon production channel with standard model (SM) particles running in the loop, annihilating DM is typically in tension with the 130 GeV gamma-ray line scenario~\cite{Asano:2012zv}. 

We consider decaying DM in this letter. Specifically, we study gravitino DM in R-parity violating (RPV) supersymmetric (SUSY) models~\cite{Takayama:2000uz,Buchmuller:2007ui}.~\footnote{Another viable decaying SUSY DM candidate, axino, has been studied in~\cite{Endo:2013si}.} With bilinear RPV operators, it is difficult to realize the gamma-ray line scenario~\cite{Buchmuller:2012rc}. The branching ratio of the radiative decay is smaller than 0.03, which is consistent with the continuum gamma-ray bound but the allowed range of parameters are constrained due to the absence of anti-proton flux in observations. We complement the previous study by considering trilinear RPV operators. With sfermion masses of $ O(\rm TeV)$, the tree-level decay rate of gravitino is suppressed and the radiative decay can explain the 130 GeV gamma-ray line. Furthermore, there is no overproduction of continuum gamma-ray and anti-proton overproduction is avoided when one considers $LLE$ RPV operators. 

The model considered here is consistent with cosmology. The lightest SUSY particle of the MSSM (MSSM-LSP) decays into gravitino or other SM particles due to RPV interactions before the big-bang nucleosynthesis (BBN) begins.
This prevents the late decay of the MSSM-LSP from spoiling the success of the BBN. We also note that the requirement of relatively heavy sfermions does not contradict with the recent discovery of the 126 GeV Higgs-like boson at the LHC~\cite{Aad:2012tfa,Chatrchyan:2012ufa}. In fact, negative results on SUSY searches at the LHC and a rather heavy Higgs boson favor SUSY models with large sfermion masses. 

The rest of the letter is organized as follows. First, we will discuss the theoretical framework of our model. Then, we discuss  general aspects of its phenomenology. Next, we elaborate how it can explain the 130 GeV gamma-ray line. Before we comment on our model and make conclusion, we study cosmological aspects of the model.   

\section{Gravitino dark matter with R-parity violation} 
\label{sec:gravitino}

\subsection{Framework} \label{sec:m}

Let us write down the relevant interaction lagrangian of gravitino:
\begin{equation}
 \begin{split}
  \mathcal L_{int} =&-\frac{i}{\sqrt{2}\,\MP}\left[ \left( D_{\mu}^*\phi^{i*}\right) \bar{\psi}_{\nu}\gamma^{\mu}\gamma^{\nu}P_L\chi^i-\left( D_{\mu}\phi^i\right) \bar{\chi}^iP_R\gamma^{\nu}\gamma^{\mu}\psi_{\nu}\right] \\
&-\frac{i}{8\MP}\,\bar{\psi}_{\mu}\left[ \gamma^{\nu},\,\gamma^{\rho}\right] \gamma^{\mu}\lambda^{(\alpha)\,a}F_{\nu\rho}^{(\alpha)\,a}+\mathcal{O}(\MP^{-2})\,. 
\label{eq:grav}
\end{split}
\end{equation}
$\phi^{i},\chi^{i},\,\psi_{\nu},\,\lambda^{(\alpha)\,a}$\,are sfermion, the corresponding fermion, gravitino and gaugino, respectively. $D_{\mu}$\,is the covariant derivative and $F_{\nu\rho}^{(\alpha)\,a}$\,is the field strength tensor. $P_L\,(P_R)$\,is the projection operator projecting onto left-handed (right-handed) spinors. All interactions are suppressed by the reduced Planck mass~$\MP\simeq2.4\times 10^{18}\,\rm GeV$\,. 

Next, we write down the superpotential related to RPV~\cite{Barbier:2004ez}. In the most general form, it is
\begin{equation}
W= \lambda_{ijk} L_{i}L_{j}{E}_{k}+\lambda_{ijk} ^{\prime }L_{i}Q_{j}{D}_{k}
+\lambda_{ijk} ^{\prime \prime }{U}_{i}{D}_{j}{D}_{k}+\mu_i L_i H_u,  \label{W_bi} 
\end{equation}
where summation among the indices $i,j,k=1,2,3$ which denote the lepton and quark generation is implicitly assumed. $L_i$, $E_i$, $Q_i$, $D_i$, $U_i$ and $H_u$ are chiral superfields of lepton doublet, lepton singlet, quark doublet, down-type quark singlet, up-type quark singlet and up-type Higgs doublet, respectively. The first three terms lead to trilinear RPV while bilinear RPV arises due to the last term. $\lambda_{ijk},\,\lambda_{ijk} ^{\prime },\,\lambda_{ijk} ^{\prime \prime }$\,are dimensionless parameters 
and $\mu_i$ is a parameter with mass dimension one.
Hereafter, we will work in the basis where $\mu_i L_i H_u$\,is rotated away from the superpotential. This is done by 
redefining $L_i$ and the down-type Higgs superfield $H_d$ as
$L_i' = L_i - \epsilon_i H_d$ and $H_d' = H_d +\epsilon_i L_i$ with $\epsilon_i \equiv \mu_i/\mu$,
where $\mu$ is the higgsino mass parameter in the MSSM superpotential $\mu H_u H_d$. Due to this redefinition, SUSY-breaking soft terms, including those corresponding to the bilinear RPV,
\begin{equation}
- \mathcal L_{soft} = (B H_u H_d + B_i\tilde{L}_iH_u + m^2_{L_iH_d}\tilde{L}_iH^*_d + {\rm h.c.}) +m^2_{H_d}|H_d|^2 +m^2_{\tilde{L}_i}|\tilde{L}_i|^2 + ...\,, 
\label{eq:soft}
\end{equation}
where $\tilde{L}_i$ is the scalar component of the chiral superfield $L_i$, undergo transformation as well. 
In the following, primes of the redefined fields and soft terms are omitted to tidy up our notations. In this basis, sneutrinos' vacuum expectation values (VEVs) are typically non-zero. By minimizing the scalar potential of sneutrino (including the SUSY-breaking soft terms), the VEVs are found to be
\begin{equation}
	\langle \tilde \nu_i\rangle = -\frac{m_{L_i H_d}^2\cos\beta + B_i\sin\beta}{m_{\tilde\nu_i}^2}v.\label{nuVEV}
\end{equation}
$\tan\beta \equiv v_u/v_d$ is the ratio of the VEVs of the up- and down-type Higgs fields. $m_{\tilde\nu_i}^2$ is the sneutrino mass and
$v \equiv \sqrt{v_u^2+v_d^2}\simeq 174$\,GeV. 

Note that trilinear RPV terms, $LLE$ and $LQD$, are also generated by the field redefinition. They are absorbed into the parameters $\lambda_{ijk}$\,and $\lambda^{\prime}_{ijk}$. In summary, we work with the superpotential
\begin{equation}
W= \lambda_{ijk} L_{i}L_{j}{E}_{k}+\lambda_{ijk} ^{\prime }L_{i}Q_{j}{D}_{k}
+\lambda_{ijk} ^{\prime \prime }{U}_{i}{D}_{j}{D}_{k}
\label{genw} 
\end{equation}
and non-zero sneutrino VEVs, which we parametrize as $\kappa_i \equiv \langle \tilde \nu_i\rangle/v $.

\subsection{Some phenomenological implications and constraints} \label{i}

In this subsection, we consider general phenomenological aspects of our framework. First, we focus on trilinear RPV couplings. Allowing both lepton and baryon number violation would lead to proton decay with a very short lifetime. Moreover, as will be discussed in detail in the next subsection, gravitino DM's leptonic decays are preferred over hadronic decays in order to explain the 130 GeV gamma-ray line. Therefore, we assume baryon number conservation by choosing $\lambda_{ijk}^{\prime \prime}=0$\,.  

Another bound on trilinear RPV couplings arises from cosmological considerations. $\lambda_{ijk}$\,and $\lambda_{ijk}^{\prime}$ has to be small enough to prevent wash-out of the baryon asymmetry before the electroweak transition. The couplings are, generically~\cite{Campbell:1990fa,Fischler:1990gn,Dreiner:1992vm,Endo:2009cv}\,,
\begin{equation}
\lambda_{ijk}\,,\lambda_{ijk}^{\prime} \lesssim \,5 \times10^{-7} \left(\frac{M_{\rm SUSY}}{1 \rm TeV}\right)^{1/2}
\label{eq:bound}
\end{equation}
where $M_{\rm SUSY}$ is the masses of squarks or sleptons.
Other bounds on trilinear RPV couplings are known to be less stringent~\cite{Barbier:2004ez}.
 
We can no longer distinguish between the lepton doublet and the up-type Higgs doublet under bilinear RPV.  Sneutrino VEVs that mix leptons with gauginos are induced. Specifically, neutrinos mix with neutralinos whereas leptons mix with charginos. Bilinear RPV's constraints can be deduced from neutrino masses generated by sneutrino VEVs. 
Masses of neutrino are $m_\nu \sim g^2 \langle \tilde \nu\rangle ^2 / m_{\tilde {B}}$, 
where $m_{\tilde {B}}$ is the mass of bino~\cite{Romao:1999up,Takayama:1999pc,Hirsch:2004he}.
For gaugino masses of $O(\rm TeV),$\,$\kappa_i = \langle \tilde \nu_i\rangle/v \lesssim 10^{-6}$ is  required by experimental bounds on neutrino masses.

In the following, we will study the scenario where trilinear RPV is dominant and bilinear RPV is negligible.~\footnote{See~\cite{Feng:2013vva,Ibe:2013jya} for models that generate small $LLE$ RPV operators, which will be important in the next section. See also~\cite{Shirai:2009fq}.} We note that even if bilinear RPV is absent at tree level at a certain energy scale, renormalization group evolution will generate the bilinear terms at some other energy scale. Unfortunately, we have not found any model that can naturally explain the smallness of bilinear RPV in the literature. On the flip side, if the 130 GeV gamma-ray line can really be interpreted as a signature of gravitino DM with trilinear RPV, it should inspire model building efforts towards a theory with such characteristic in the future.

\subsection{Trilinear R-parity violation-dominant scenario}  \label{sec:tri}

Gravitino undergoes three-body decay via trilinear couplings at the tree level. As will be explained in the following, we will consistently be working with the $LLE$ RPV operators. For leptonic decay with an intermediate mass $m_{\tilde \tau_R}$, the decay rate is~\cite{Buchmuller:2007ui}
\begin{equation}
\label{eq:tree}
\Gamma(\psi_{3/2} \to \bar\tau \nu_i e_j) \simeq \frac{|\lambda_{ij3}|^2}{90(32)^2\pi^3}\frac{m_{3/2}^7}{\MP^2 m_{\tilde \tau_R}^4}.
\end{equation}
The full analytical result has been worked out in~\cite{Moreau:2001sr}. Gravitino undergoes one-loop radiative decay as well~\cite{Lola:2007rw}. The decay rate scales as
\begin{equation}
\Gamma(\psi_{3/2} \to \gamma \nu_i) \sim \frac{\alpha \lambda_{ijj}^2 m_{3/2} m_{j}^2}{\MP^2}
\end{equation}
when the sfermion masses are large compared to gravitino and lepton masses. $\alpha$ is the fine structure constant. The radiative decay rate is proportional to the fermion mass. Therefore, RPV couplings involving the third generation of fermion gives the largest contribution. We also note that the radiative decay is approximately independent of the mass of sfermion running in the loop of the decay amplitude. It can be understood from the amplitude of the interaction that involves the first two terms of Eq.\,(\ref{eq:grav}). These terms carry a derivative of the sfermion field that brings a momentum flow proportional to the largest loop-mass to the vertex. It balances out the contribution of the sfermion mass from the sfermion propagator.~\footnote{See~\cite{Lola:2007rw} for a detailed description of the radiative decay amplitude.} 

Gravitino also decays into gauge bosons via $LLE$ RPV couplings. Similar to the radiative decay channel, in the large sfermion masses limit, the decay rate of these channels are approximately independent of sfermion masses.  Ignoring numerical prefactors, the decay rate for $\psi_{3/2} \to Z \nu_i$ scales as
 \begin{equation}
 \label{eq:znu}
\Gamma(\psi_{3/2} \to Z \nu_i) \sim \frac{\alpha \lambda_{ijj}^2 m_{3/2} m_{j}^2}{{\rm sin}^2 \theta_W\MP^2}\frac{m_{3/2}^2-m_Z^2}{m_{3/2}^2},
\end{equation}
where  $\theta_W$ is the Weinberg angle. For $\psi_{3/2} \to W^+ l_i^-$, one simply replaces $(m_{3/2}^2-m_Z^2)/m_{3/2}^2$ of Eq.\,(\ref{eq:znu}) with $[(m_{3/2}^2-(m_W-m_{l_i})^2)(m_{3/2}^2-(m_W+m_{l_i})^2)]^{1/2}/m_{3/2}^2$. The full analysis of these channels has been worked out in~\cite{Batzing:1}. 

\begin{figure}
\begin{center}
\includegraphics[scale=1.2]{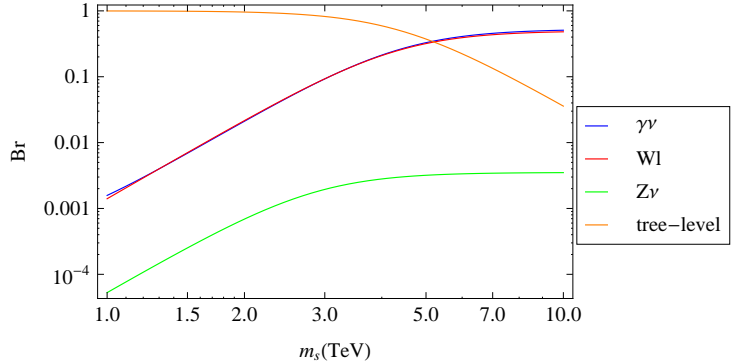}
\caption{Dependence of branching ratios of gravitino decay via $\lambda_{133}$ on the intermediate sfermion mass $m_s$. Here, all sfermions are assumed to take the value $m_s$. We have used Eq.\,(\ref{eq:tree}) for the tree-level decay. Decay rates of radiative decay and decays into gauge bosons are taken from~\cite{Lola:2007rw}  and~\cite{Batzing:1} respectively. Note that the plots of radiative channel overlap with  $\psi_{3/2} \to Wl$ channel's. The mass of the gravitino is $m_{3/2} \simeq 260 \rm GeV$.}
\label{fig:fig}
\end{center}
\end{figure}

The dependence of these branching ratios on the sfermion mass is shown in Fig. \ref{fig:fig}. As expected, the branching ratio of the tree-level decay drops along with the increase of sfermion mass. One-loop decay channels become important in the heavy sfermions limit. As can be seen in Fig. \ref{fig:fig}, by taking $m_{3/2} \simeq 260 \rm GeV$ and increasing the mass of sfermions, one can get the branching ratio $\Gamma(\psi_{3/2} \to \gamma \nu)$ that is large enough to explain the Fermi line. 

The apparent 130 GeV gamma-ray line can be explained by a decaying DM of lifetime $\tau_{\rm DM}$\,that satisfies
 \begin{equation}
 \tau_{\rm DM}/{\rm Br}({\rm DM}\to \gamma\nu) = (1-3)\times 10^{28}\, \rm sec  
\label{eq:rate}
 \end{equation}
 and branching ratio ${\rm Br}({\rm DM}\to \gamma\nu)\gtrsim 0.01$~\cite{Buchmuller:2012rc}. In this region of parameter, astrophysical constraints from diffuse gamma-ray~\cite{Ackermann:2012rg} and neutrino spectra~\cite{Bomark:2009zm,Covi:2009xn} are also satisfied. As an illustration of our model, we choose the sfermion masses to be $m_s\simeq 3\,\rm TeV$. The branching ratio of the radiative decay is ${\rm Br}({\rm DM}\to \gamma\nu)\simeq 0.1$. In order to conciliate with Eq.\,(\ref{eq:rate}),  the RPV coupling are needed to be $\lambda \equiv \lambda_{133} \simeq 6 \times 10^{-7}$.
 
We note that a large range of parameters (DM lifetime and branching ratio) is excluded by the PAMELA anti-proton data for gravitino DM with bilinear RPV~\cite{Buchmuller:2012rc}. We do not have such concern for gravitino with $LLE$ RPV operators when the tree-level decay channel is dominant. This is because this channel does not produce anti-proton. Even when one-loop decays are dominant, the anti-proton bound is much relaxed as the radiative decay channel gives large, if not the largest branching ratio. 

We now discuss several cosmological constraints on our model. Under trilinear RPV, the MSSM-LSP, which we assume to be the bino, may decompose into SM particles via tree-level decay. The decay rate is
\begin{equation}
\Gamma_{\tilde{B} \to SM} = \frac{5 \lambda^2 \alpha}{16 \pi^2 {\rm cos^2}\theta_{W}}\,m_{\tilde{B}} \psi(m_{s}/m_{\tilde{B}}).
\end{equation}
We have assumed that the masses of left-handed and right-handed stau are degenerate, i.e. $m_s \equiv m_{\tilde{\tau_R}}=m_{\tilde{\tau_L}}.$\,The function $\psi(y)$ is defined as
\begin{equation}
\psi(y)=\int^{1/2}_0 dx \frac{x^2(1-2x)}{(1-2x-y^2)^2}.
\end{equation}
For $m_{s}/m_{\tilde{B}} \lesssim 10$ and $m_{\tilde{B}}\sim 1 \rm TeV$,\,$\lambda$ has to be greater than $10^{-9}$ so that $\Gamma_{\tilde{B} \to SM}^{-1} \lesssim 1 \rm sec$. We see that in the region of parameter of interest, BBN is unaffected as bino decays much earlier than $ 1\,\rm sec$. 

MSSM-LSP decays into gravitino as well. The decay rate is~\cite{Covi:2009bk}
\begin{equation}
\Gamma^{-1}_{\tilde{B} \to \psi_{3/2}} \simeq 5 \times 10^4 \rm sec \left(\frac{m_{\tilde{B}}}{1 \rm TeV}\right)^{-5} \left(\frac{m_{3/2}}{260 \rm GeV}\right)^2.
\end{equation} 
Bino that decays into gravitino contribute to the relic abundance of gravitino. However, since $\Gamma_{\tilde{B} \to SM}^{-1}\ll \Gamma_{\tilde{B} \to \psi_{3/2}}^{-1}$,  only an insignificantly small fraction of bino decays into gravitino. Hence, bino's contribution to the gravitino relic abundance is negligible.

The thermal relic abundance of gravitino is~\cite{Moroi:1993mb,Bolz:2000fu,Pradler:2006qh,Rychkov:2007uq}
 \begin{equation}
\Omega_{3/2}h^2\simeq 0.1\left( \frac{T_R}{10^{10}\,\rm{GeV}}\right) \left( \frac{m_{3/2}}{260\,\rm{GeV}}\right)^{-1} \left( \frac{m_{\tilde{g}}}{1\,\rm{TeV}}\right) ^2, 
\end{equation}
where $T_R$ is the reheating temperature and $m_{\tilde{g}}$ is the gluino mass. Since thermal leptogenesis requires $T_R\gtrsim 10^9$~\cite{Buchmuller:2004nz}, our scenario is consistent with thermal leptogenesis for  $m_{\tilde{g}}\sim O(\rm TeV)$.

\section{Discussion and Conclusion} 
\label{sec:conc}
Several comments are in order before we conclude. The morphology of the observed gamma-ray line excess favors annihilating DM but decaying DM is acceptable as well~\cite{Buchmuller:2012rc,Park:2012xq}. The Einasto and NFW profiles are compatible with annihilating DM but a strongly contracted profile is needed in the case of decaying DM. More data is required before one confirms or rules out the possibility of explaining the gamma-ray line with decaying DM. 

We now briefly discuss the prospect of detection in collider. The lifetime of the MSSM-LSP is around $10^{-6}-10^{-4}$ sec and thus if produced, it will decay outside of the detector. Events with missing energy can be recognized as the collider signature. 

In conclusion, we have shown that gravitino dark matter with trilinear R-parity violation is capable of explaining the 130 GeV gamma-ray line. Models with the $LLE$ R-parity violating coupling are especially advantageous since there is no overproduction of anti-proton flux, in contrast with the bilinear R-parity violating scenario. Other astrophysical constraints are also satisfied. Furthermore, our model is consistent with cosmology (big-bang nucleosynthesis and thermal leptogenesis). The requirement of sfermion masses of $O(\rm TeV)$ is well-motivated by the 126 GeV Higgs boson and negative searches for supersymmetric particles.

\section*{Acknowledgment}
The author is grateful to K. Hamaguchi for various discussions and suggestions. The author would also like to thank K. Ohmori for providing technical assistance.



\end{document}